\documentclass{mem_l}
\usepackage{natbib}\usepackage{txfonts}\usepackage{balance}
\usepackage{graphicx}
\usepackage[a4paper]{hyperref}
\idline{75}{282}
\begin{document}
\def\teff{$T\rm_{eff }$}
\def\kms{$\mathrm {km s}^{-1}$}

\title{
Kurucz's codes under GNU-Linux
}

\author{
L. Sbordone \inst{1,2} 
          }

  \offprints{L. Sbordone}

\institute{
Istituto Nazionale di Astrofisica --
Osservatorio Astronomico di Trieste, Via Tiepolo 11,
I-34131 Trieste, Italy
\and
Istituto Nazionale di Astrofisica -- Osservatorio Astronomico di Roma
via di Frascati 33, I-00040 Monte Porzio Catone, Italy
\email{sbordone@oa-roma.inaf.it}
}

\authorrunning{Sbordone}

\titlerunning{Kurucz's codes under Linux}

\abstract{We report on the latest version of our GNU-Linux port of the ATLAS -- SYNTHE -- WIDTH suite of codes for the stellar atmosphere modeling. The latest version (8.1 at the time of the workshop) of the Intel Fortran Compiler allowed for a significantly better backward compatibility with the VMS version of the code, thus allowing us to remove almost all the modifications we initially introduced in order to compile the code under IFC.
We now provide ported versions both of ATLAS 9 (the ODF version of ATLAS) and ATLAS 12 (the opacity sampling version).
A comprehensive website has been created to host the ported codes, along with a growing body of documentation and resources. Also, three mailing lists have been created at the university of Ljubljana in order to cover general usage, code development and ATLAS-related announcements.
}
\maketitle{}

\section{Introduction}

In \citet[][henceforth Paper 1]{sbordone04atlas} we reported the availability of a GNU-Linux port of the suite of codes by R. L. Kurucz for stellar atmosphere modeling, spectral synthesis and abundance analysis \citep[henceforth ATLAS suite,][]{kurucz93}. The ATLAS suite has been, along the years, one of the main tools employed in the synthesis and analysis of stellar atmospheres and spectra, but it is only actively maintained by Kurucz himself under VMS, an operating system now far less widespread than Linux. The computing power of contemporary personal computers makes it possible to effectively use the ATLAS suite on a low-end laptop, thus making the porting highly useful. At the same time, GNU-Linux is being massively adopted for high performance computing applications, a trend also due to the possibility it gives to build powerful clusters by using mainstream PC hardware (e. g. Beowulf systems).

\section{The Linux port}

We aimed to produce a port that remained as near as possible to the VMS version, both to assure two strongly compatible versions, and to limit the rewriting effort needed to compile the suite under Linux. In fact, the major challenge we faced was related to the presence of large amounts of obsolescent Fortran constructs in the ATLAS suite, which was developed along the last 40 years. We thus needed a compiler with a powerful set of backward-compatibility options, which led us to discard the standard GNU-fortran compiler G77. Instead we adopted the proprietary Intel Fortran Compiler (IFC), licensed free of charge for research applications under Linux. IFC is also an extremely powerful compiler for scientific applications, and allows for strong optimization both on 32 and 64 bit architectures, producing significant faster executables than G77.

The use of IFC allowed us to port the suite by introducing a small set of changes, detailed in Paper 1. Most of the porting was carried out using IFC version 7.1, although, at the time of the publication of Paper 1, IFC 8.0 was already available and was used for the performance benchmarking. In the following months, mainly due to the efforts of F. Castelli, we discovered that, with the newer versions of IFC (in the meantime version 8.1 had become available), many of the modifications we introduced in the first port were no longer necessary. Among other things, IFC now interprets correctly all the used Hollerith statements and accepts unnamed {\tt OPEN} statements (e. g. {\tt OPEN(7,STATUS="OLD")} looks for a file called {\tt fort.7}). As a consequence, we updated the porting of the code and translated the latest VMS version available on R. L. Kurucz website\footnote{{\tt http://kurucz.harvard.edu/}}. Of the changes in ATLAS 9 cited in Paper 1 only three remained in the latest version of the port:
\begin{itemize}
\item A single {\tt COMMON} statement shared in the Kurucz version by the three subroutines {\tt JOSH}, {\tt BLOCKJ}, and {\tt BLOCKH} was splitted in two different {\tt COMMON} statements, the one shared by {\tt JOSH} and {\tt BLOCKJ}, the other shared by {\tt JOSH} and {\tt BLOCKH}. 
\item All the calls to the {\tt ABORT} function have been substituted by calls to {\tt EXIT}.
\item All the calls to {\tt BEGTIME} and {\tt ENDTIME} were dropped.
\end{itemize}

\section{Available resources and web site}
The port described in Paper 1 was available upon request, while we were working setting up a more comprehensive web site. The site went online just before the present Workshop\footnote{{\tt http://wwwuser.oat.ts.astro.it/atmos/}}.

The ATLAS suite port is currently freely available from the {\em Download} section of the web site, where it is divided into various archives:
\begin{description}
\item[{\bf src.tar.bz2}] source codes for all the ported programs. Included are ATLAS 9 (using Opacity Distribution Functions, henceforth ODF), ATLAS 12 (opacity sampling version), the suite of programs globally known as SYNTHE, for the production of synthetic stellar spectra, WIDTH (abundance determination from equivalent widths). Included are also some general use utilities from R. L. Kurucz's web site, e.g. for synthetic spectra instrumental ({\tt broaden}) and rotational ({\tt rotate}) broadening.
\item[{\bf utilities.tar.bz2}] Source codes for (contributed) utilities. It essentially includes all the programs not ported from R. L. Kurucz repository but written by someone else. It is going to include any other code that will be considered of general interest.
\item[{\bf bin.tar.bz2}] Precompiled binaries of both Kurucz's and contributed codes. They are statically compiled with IFC (version 8.1 at the time of the Workshop), so they are supposed to run on any Linux distribution meeting the IFC system requirements. During the Workshop's practical sessions the precompiled binaries were installed on a variety of Linux distributions, showing no problems except in a couple of Mandriva systems. We did not further investigate the origin of the issue, a local recompilation fixed the problem.
\item[{\bf lines.tar.bz2}] ASCII linelists from R. L. Kurucz website, used mainly by SYNTHE.
\item[{\bf lines\_at12.tar.bz2}] This 1 GB file includes the linelists needed by the Opacity Sampling version of ATLAS. Users planning not to use ATLAS 12 may avoid downloading it.
\item[{\bf ODF.tar.bz2}] Contains the Opacity Distri-bution Functions needed by ATLAS 9. We included a representative set of ODFs of the ``new'' type \citep[see][]{castelli03}. ``Old'' type ODFs can be downloaded from R. L. Kurucz's site.
\item[{\bf molecules.tar.bz2}]molecular linelists, in ASCII format, from R. L. Kurucz web page, used by SYNTHE.
\item[{\bf scripts.tar.bz2}] Sample scripts for launching the various programs. They are very basic shell scripts which we kept deliberately simple, to avoid confusing the new user. Of course much more complicated or automated scripts can be produced.
\item[{\bf documentation.tar.bz2}] For user's convenience, all the documentation available in the site has been made available as a single archive.
\end{description}

The ODFs used by ATLAS 9 are produced by means of the DFSYNTHE code which is not presently included in the described port. Nevertheless, it has been ported to Linux by F. Castelli and can be downloaded from her web page\footnote{{\tt http://wwwuser.oat.ts.astro.it/ castelli/sources.html}}, if particular ODFs are needed by the user. For more details on DFSYNTHE and ATLAS 12 usage see F. Castelli and R. L. Kurucz papers, in these same proceedings.

As above stated, we are trying to collect and improve the available documentation on the ATLAS suite structure and usage. The {\em Documentation} section of the website contains (or links to) the material we were able to find, or that was contributed. A basic html ``cookbook'' is also provided which describes the use of ATLAS 9, WIDTH and SYNTHE under Linux with the basic input scripts and input files that are provided in the scripts archive. We expect documentation to grow in time thanks to the contribution of the community of the ATLAS users.

During the Workshop, the need has emerged for better interaction within the ATLAS users community. To accomplish this, three mailing lists have been created, and are maintained, by U. Jauregi at the Ljubljana University, Slovenia. They are aimed to deal with general user needs and discussions ({\tt kurucz-discuss}), important announcements such as conferences and workshops ({\tt kurucz-announce}), and more specific code development discussions ({\tt kurucz-devel}). Their home pages, useful also to access the mailing list archives, can be reached from the {\em Links} section of our website, along with some other links of interest.

\section{Future work and conclusions}
The availability of the ATLAS suite under GNU-Linux allows a much larger user base to directly access the codes. Under Linux (and generally UNIX) shell scripting and languages such as Perl or Python allow to produce a great variety of applications, from graphical front-ends to the codes, to web interfaces, to ``model factories'' to automatically produce huge grids of models and synthetic spectra for the purpose e.g. of creating integrated light synthetic spectra. Given the availability of ATLAS 12 and DFSYNTHE, moreover, arbitrary chemical compositions can be used easily in creating such grids. Given the high speed of the ported codes \citep[see][for a comparison with the speed on VMS workstations]{sbordone04atlas}, and the availability of cheap and scalable Linux clusters such as the Beowulf systems, the time and effort needed to produce even huge grids can be greatly reduced.

In creating the web site and the mailing lists, our goal was not only to make our port available to the community, but also to provide a repository for related scripts, codes and documentation that the community will be willing to share. The need for a larger and more comprehensive documentation seems to be the most urgent, along with the need of support new users. We hope to to be able to support these and further needs that could eventually become manifest in the future.

\bibliographystyle{aa}

\end{document}